\documentclass[aps,twocolumn,prd,nofootinbib]{revtex4}
\usepackage{amsmath}
\usepackage{graphicx}
\usepackage{dcolumn}
\usepackage{bm}
\usepackage{amssymb}
\usepackage{latexsym}

\begin{document}

\title{Cosmic Duality in Quintom Universe}

\author{Yi-Fu Cai$^a$, Hong Li$^{a,b}$, Yun-Song Piao$^{c}$ and Xinmin Zhang$^a$}
\affiliation{${}^a$Institute of High Energy Physics, Chinese Academy
of Sciences, P.O. Box 918-4, Beijing 100049, P. R. China}
\affiliation{${}^b$Department of Astronomy, Peking University,
Beijing 100871, P. R. China} \affiliation{${}^c$College of Physical
Sciences, Graduate School of Chinese Academy of Sciences, Beijing
100049, China}

\begin{abstract}

 In this paper we study the duality in two-field Quintom models of
 Dark Energy. We find that an expanding universe dominated by
 Quintom-A field is dual to a contracting universe with Quintom-B field.

\end{abstract}

\maketitle

The recent data from type Ia supernovae and cosmic microwave
background (CMB) radiation and so
on\cite{1998snia,Riess,Spergel,Seljak} have provided strong
evidences for a spatially flat and accelerated expanding universe
at the present time. In the context of Friedmann-Robertson-Walker
cosmology, this acceleration is attributed to the domination of a
component, dubbed dark energy (DE).
 Theoretically, the simplest candidate for DE is a
small positive cosmological constant, but it suffers from the
difficulties associated with the fine tuning and the coincidence
problems. So many physicists are attracted by the idea that dark
energy is due to a dynamical component, such as the Quintessence,
K-essence, Phantom or Quintom.

With the accumulated observational data (e.g. SNIa, Wilkinson
Microwave Anisotropy Probe observations(WMAP), galaxy
clustering(SDSS) and so on) it becomes possible in the recent
years to probe the recent and even early behavior of DE by using
some parameterizations for its equation of state (EOS) and
constrain the models of dark energy. Especially, the new released
3-year WMAP data (WMAP3)\cite{3wmap} have given so far the most
precise probe on the Cosmic Microwave Background (CMB) Radiations.
Although the recent fits to the data in combination of the WMAP3
with the other cosmological observational data show remarkably the
consistence of the cosmological constant, it is worth noting that
a class of dynamical models with equation of state across $-1$
{\it Quintom} is mildly favored \cite{Zhao:2006bt,Wang:2006ts}. In
the literature there have been a lot of studies on this class of
models \cite{Quintom2,Quintom3,Oscillating
quintom,weihao:hes,single:li,zhangxin:stafi,cai:brane,hli:sn-nu,fengbo,Wei,zhang-qiu,
zk06,zhangxin:holo,HSzhang05,Sadjadi06}. The similar work applied
in scalar-tensor theory is also studied in Ref.\cite{Elizalde}. In
the context of string theory it has been shown that the crossing
$w=-1$ can be realized as low energy limit of the dynamical
behavior of slow-rolling tachyon on a non-BPS
D3-brane\cite{stringorigin}.

 The simplest Quintom model consists of two scalar fields, one is Quintessence-like and another
is Phantom-like\cite{Quintom1,Quintom3}. The Quintom models differ
from the Quintessence or Phantom in the determination of the
evolution and fate of the universe, for example, as shown in
Ref.\cite{Oscillating quintom}, Quintom model can give rise to a new
scenario of the evolution of the universe. In this model with an
oscillating equation of state the early inflation and current
acceleration are unified. Moreover, with Quintom dark energy it
might be possible to avoid the singularities of the universe like
big rip or big crunch. In terms of the two-field Quintom models, it
has been shown recently that there exists two category \cite{zk06}:
one is Quintom-A where at early time the Quintessence dominates with
$w>-1$ and lately the Phantom dominates with $w<-1$; and the other
is Quintom-B for which the equation of state is arranged to change
from below -1 to above -1. It is trivial to realize Quintom-A in
model building; but to achieve Quintom-B, one may need not only to
add more degrees of freedom, but also to fine tune the potentials of
Quintom fields(see Ref.\cite{zk06}), or to construct non-canonical
field with high derivative terms(see Ref.\cite{single:li}), or even
to add interactive terms which provide a transition from
Phantom-like to Quintessence-like(see Ref.\cite{Quintom2}). Both
Quintom-A and Quintom-B are consistent with the current
observational data.

To understand the possible connections among the dark energy models,
it is useful to study the cosmic duality. The dualities in field
theory and string theory have been widely studied and in fact it
predicts a lot of interesting phenomena\cite{Polchinski}. The
authors of Ref.\cite{Chimento0,Chimento1} have considered a possible
transformation with the Hubble parameter and studied the relevant
issues with the cosmic duality\cite{Veneziano,Lidsey}. Specifically
Ref.\cite{Chimento2} has shown a link between a standard cosmology
with quintessence matter and a contracting cosmology with phantom.
Later on this duality has been generalized into studies with more
complicated dark energy models and it has been shown to exist for
these various dark energy models\cite{Chimento3,Chimento4,
Dabrowski1,Chimento5}. In Ref.\cite{Dabrowski2,Dabrowski3} the
authors have studied this type of duality and its connection to the
fates of the universe. In Ref.\cite{Singh}, the author has also
discussed the possibility of realizing this cosmic duality in the
braneworlds. In all of these studies, the dark energy models will
not be able to give $w$ crossing -1. Further, with phantom dominant
the universe will reach a $big$ $rip$ or $big$ $sudden$ in the
future\cite{fate1,GPZ}, or expands forever approaching to a de
Sitter, so a contracting universe is unstable.
 In this paper, we study the implications of the cosmic duality
in the Quintom models of dark energy. By studying the
behavior of the equation of state we find a dual of the
Quintom-A to the Quintom-B.

To begin with the discussion, we consider a model where the universe
is filled with Quintom dark energy and we neglect the contribution
of the components of matter and radiation. The quintom model we will
study in this paper consists of two fields, one being
quintessence-like, another phantom-like with the
 lagrangian given by
\begin{eqnarray}
\label{lagrangian} {\cal
L}=\frac{1}{2}\partial_\mu\phi_1\partial^\mu\phi_1-\frac{1}{2}\partial_\mu
  \phi_2\partial^\mu\phi_2-V_1(\phi_1)-V_2(\phi_2),
\end{eqnarray}
where $V_1$ and $V_2$ are the potential terms and we have
neglected the interaction between the $\phi_1$ and $\phi_2$ for
simplicity of the discussion.

In the framework of FRW cosmology, the Einstein equations are
\begin{eqnarray}
\label{Einstein1}
3H^2=\frac{1}{2}{\dot\phi_1}^2-\frac{1}{2}{\dot\phi_2}^2+V_1+V_2,\\
\label{Einstein2}
\ddot\phi_1+3H\dot\phi_1+\frac{dV_1}{d\phi_1}=0,\\
\label{Einstein3} \ddot\phi_2+3H\dot\phi_2-\frac{dV_2}{d\phi_2}=0.
\end{eqnarray}
Obviously we can construct a form-invariant transformation by
defining a group of new quantities $\bar H$, $\bar\rho$, $\bar p$
and $\bar w$ which keep the Einstein equations invariant. Following
the similar work of Ref.\cite{Chimento2} there is a form-invariant
transformation as follow:
\begin{eqnarray}
\bar\rho&=&\bar\rho(\rho),\\
\bar{H}&=&-(\frac{\bar\rho}{\rho})^{\frac{1}{2}}H.
\end{eqnarray}
Under this transformation, we obtain the corresponding changes for the
pressure $p$ and the equation of state $w$,
\begin{eqnarray}
\label{ptrans}\bar{p}&=&-\bar\rho-(\frac{\rho}{\bar\rho})^{\frac{1}{2}}(\rho+p)\frac{d\bar\rho}{d\rho},\\
\label{wtrans}\bar{w}&=&-1-(\frac{\rho}{\bar\rho})^{\frac{3}{2}}\frac{d\bar\rho}{d\rho}(1+w).
\end{eqnarray}
From eqs. (\ref{ptrans}) and (\ref{wtrans}) one can see that for a
positive $\frac{d\bar\rho}{d\rho}$, one would be able to establish a
connection between the Quintom-A and Quintom-B of the Quintom model
of the dark energy.

For the specific model of quintom dark energy we consider in this
paper (\ref{lagrangian}) the energy density and the pressure of
system are given by
\begin{eqnarray}
\rho&=&\frac{1}{2}{\dot\phi_1}^2-\frac{1}{2}{\dot\phi_2}^2+V_1+V_2,\\
p&=&\frac{1}{2}{\dot\phi_1}^2-\frac{1}{2}{\dot\phi_2}^2-V_1-V_2.
\end{eqnarray}

Taking $\bar\rho=\rho$ in (\ref{ptrans}) and (\ref{wtrans}) as an
example of detailed discussion without loss of the generality of the
physical conclusion and information, we can obtain the dual
transformation
\begin{eqnarray}
\bar H&=&-H,\\
\bar p&=&-2\rho-p,\\
\bar w&=&-2-w.
\end{eqnarray}
Consequently, the dual form of the Quintom dark energy to
(\ref{lagrangian}) is given by
\begin{eqnarray}\label{Transform of L}
\bar{\cal
L}&=&{\bar T} - {\bar V} \nonumber\\
&=&\frac{1}{2}\partial_\mu\phi_2\partial^\mu\phi_2-\frac{1}{2}\partial_\mu
  \phi_1\partial^\mu\phi_1\nonumber\\
  &&-\delta{\cal L}_1(\phi_1)-\delta{\cal L}_2(\phi_2),
\end{eqnarray}
where ${\bar T}$ and $\bar V$ denote for the kinetic and
potential energy terms of dual form, and $\delta{\cal L}_1$ and
$\delta{\cal L}_2$ are:
\begin{eqnarray}
\delta{\cal L}_1&=&V_1+{\dot\phi_1}^2,\\
\delta{\cal L}_2&=&V_2-{\dot\phi_2}^2.
\end{eqnarray}
Comparing the quintom model in (\ref{lagrangian}) and the dual form
of the model in (\ref{Transform of L}), one can see that with the
dual transformation,
 if the original lagrangian is for a Quintom-A model the dual
one is for Quintom-B model, or vice versa. With this duality,
one expects a general connection among different fates
of the
universe, and it might be possible that the early universe be linked
to other
epochs of the universe.

For a detailed discussion on the duality connecting Quintom-A and
Quintom-B in our note, we take a special form of the potentials in
the unit of planck mass
\begin{eqnarray}
\label{V_1}V_1&\propto&-3\sqrt{2}\phi_1+2e^{-\sqrt{2}\phi_1},\\
\label{V_2}V_2&\propto&\frac{3}{2}{\phi_2}^2+4\phi_2.
\end{eqnarray}
Now we study its semi-analytic solution and discuss more details of
this duality.

Solving explicitly the Einstein equation (\ref{Einstein1}) and the
equations of motion for two scalar fields
(\ref{Einstein2},\ref{Einstein3}) together, we will study
specifically the two periods of the universe evolution with this
quintom model. For the first period that $|t|\ll1$, we take the
initial conditions by fixing $\phi_1 \rightarrow -\infty$ and
$\phi_2 \rightarrow 0$. With these initial conditions one can see
that the dominant component in the energy density of the quintom
model is the exponential term of $V_1$ in eq.(\ref{V_1}), namely the
contribution from the Phantom potential in eq.(\ref{V_2}) and the
linear part of Quintessence potential in eq.(\ref{V_1}) can be
neglected. Therefore the evolution of the universe behaves like the
one dominant by Quintessence component $\phi_1$ and the universe is
evolving with the approximate solution given by explicitly:
\begin{eqnarray}\label{result1}
\phi_1\sim\sqrt{2}\ln{|t|}, \phi_2\sim\frac{1}{2}t^2,
H\sim\frac{1}{t}.
\end{eqnarray}
and thus one can see that the scale factor in this period would
variate with respect to time of form $a\propto\pm t$ in which the
signal is determined by the positive definite form of the scale
factor. Therefore the scale factor here is corresponding to the big
bang or big crunch of Quintessence-dominant universe.

The dual form of the solution above is a description of one universe
dominant by a Phantom component with a lagrangian given by
eq.(\ref{Transform of L}) and
\begin{eqnarray}
\delta{\cal L}_1+\delta{\cal L}_2&=&{\dot\phi_1}^2-{\dot\phi_2}^2+V_1+V_2\nonumber\\
&\sim&-3\sqrt{2}\phi_1+4e^{-\sqrt{2}\phi_1}+2\phi_2+\frac{3}{2}{\phi_2}^2,
\end{eqnarray}
correspondingly the dual hubble parameter is of form
$\bar{H}\sim-\frac{1}{t}$ and the scale factor is of form
$a\propto\pm\frac{1}{t}$. Accordingly, the scale factor of the dual
form is tending forward to infinity in the beginning or the end of
universe. From what we have investigated in the above, one can see
that, for the positive branch there is a duality between an
expanding universe with initial singularity at $t=0^+$ and a
contracting one that begins with an infinite scale factor at
$t=0^+$. However, for the negative branch there is a duality between
a contracting universe ending in a big crunch at $t=0^-$ and an
expanding one that ends in a final big rip at $t=0^-$. The latter is
justly dominated by a phantom component. Besides, in general with
phantom dominant the contracting solution is not stable, because the
Phantom universe will run into $big$ $rip$, $big$ $sudden$ or
expands forever approaching to a de Sitter, but will not be able to
stay in the contracting phase forever\cite{fate1,GPZ}. This problem,
however, can be avoided in Quintom model since in the dual universe
with Quintom-B dark energy the increase of kinetic energy of Phantom
during the contraction can be set off by that of Quintessence at
late time.

For the case that ($|t|\gg1$), Phantom component in the quintom
model (1) will dominant and the universe will expand. For the
specific potentials we consider in eqs. (\ref{V_1}) and (\ref{V_2})
the mass term in $V_2$ will gradually play an important role in the
the evolution of dark energy. With the analysis similar to above we
have
\begin{eqnarray}
\phi_1\sim\sqrt{2}\ln{|t|}, \phi_2\sim\sqrt{2}t, H\sim t,
\end{eqnarray}
where we note that the scale factor is
$a\propto\exp(\frac{t^2}{2})$. Consequently, the scale factor would
increase forward to the infinity rapidly for the positive branch
while start from an infinity for the negative branch.

Then the transformed lagrangian now is (\ref{Transform of L}) with
\begin{eqnarray}
\delta{\cal
L}_1+\delta{\cal
L}_2
\sim
-3\sqrt{2}\phi_1+4e^{-\sqrt{2}\phi_1}+4\phi_2+\frac{3}{2}{\phi_2}^2-2.
\end{eqnarray}
The universe is evolving with the hubble parameter like
$\bar{H}\sim{-t}$, and the scale factor like
$a\propto\exp(-\frac{t^2}{2})$ which is close to singularity related
to the origin and the fate of universe. Finally, the component which
dominates the evolution of the universe is $\phi_2$, the scalar
field resembling the quintessence in (\ref{Transform of L}).
Consequently we obtain the conclusion that, for the positive branch,
there is a dual relation between an expanding universe with a fate
of expanding for ever with $t\rightarrow+\infty$ and a contracting
universe with a destiny of shrinking for ever with
$t\rightarrow+\infty$; meanwhile, for the negative one, there is a
duality connecting a contracting universe starting from near
infinity with $t\rightarrow-\infty$ and a expanding universe origin
from infinity with $t\rightarrow-\infty$.

\begin{figure}[htbp]
\begin{center}
\includegraphics[scale=0.8]{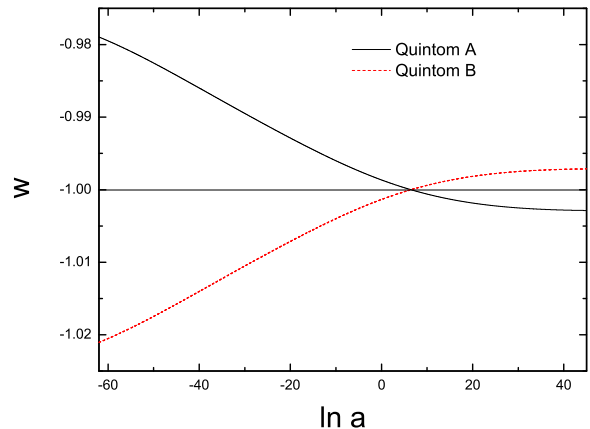}
\caption{Plot of the equation of state w of the Quintom model and its dual
as a
function of the scale factor $\ln a$ for
$V=-3\sqrt{2}M^3\phi_1+2M^4e^{-\frac{\sqrt{2}\phi_1}{M}}+\frac{3}{2}M^2{\phi_2}^2+4M^3\phi_2$
and $M$ is the planck mass. \label{Fig:dual S}}
\end{center}
\end{figure}

Having present the analytical arguments for the duality between the
quintom A and quintom B model of the dark energy we study in detail
the numerical solutions. In Fig.\ref{Fig:dual S} we plot the
evolution of the equation of state of the quintom model and its
dual. One can see from this figure that under the framework of the
duality studied in this paper, the equation of state of the quintom
model and its dual are symmetric around $w = -1$. Accordingly, in
this case the Quintom A model of dark energy is dual to a Quintom B
model rigorously, which supports for our analytical arguments above.

\begin{figure}[htbp]
\includegraphics[scale=0.8]{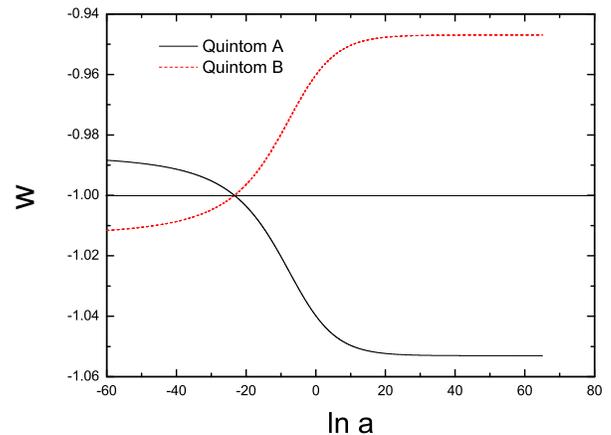}
\caption{Plot of the w of the Quintom model and its dual as a
function of the scale factor $\ln a$ for
$V=V_0(e^{-\frac{\phi_1}{M}}+e^{-\frac{2\phi_2}{M}})$ and $M$ is
the planck mass. \label{Fig:dualexp}}
\end{figure}

\begin{figure}[htbp]
\includegraphics[scale=0.8]{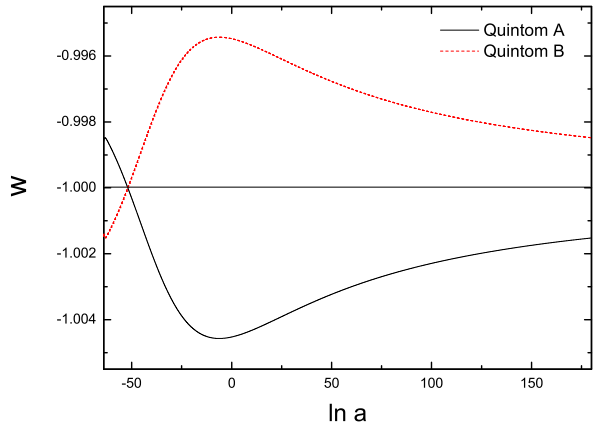}
\caption{Plot of the w of the Quintom model and its dual as a
function of the scale factor $\ln a$ for
$V={m_1}^2{\phi_1}^2+{m_2}^2{\phi_2}^2$ where $m_1$ corresponds to the
Quintessence component mass and $m_2$ the Phantom component mass.
\label{Fig:dualmass}}
\end{figure}

In Fig.\ref{Fig:dualexp} we take the potentials $V_1$ and $V_2$ to
be exponential and one can see that the equation of state for
quintom A approches to a fixed value which corresponds to the
attractor solution of this type of model\cite{Quintom2}. Through the
duality one can see that there exists a corresponding attractor of
the Quintom B model dual to the former one. In
Fig.\ref{Fig:dualmass}, we provide another examples for the duality.


In summary, we have studied the cosmic duality in the quintom
models. In general we have shown the Quintom model has its dual
partner, specifically the Quintom A model is dual to the Quintom B.
These two models describe two different behaviour of the universe
evolution with one in the expanding phase and another in the
contracting depending on the initial conditions we choose. The
cosmic duality which connects the two totally different scenarios of
universe evolution keeps the energy density of the universe
unchanged, but transforms the hubble parameter. With this duality
and keeping Einstein equations form-invariant, we have found that
Quintom A model is dual to Quintom B. As is know that with different
type of dark energy, the fate of the universe will be different. Our
study in this paper helps understand the properties of various dark
energy models and their connections to the evolution and the fate of
the universe.

This work is supported in part by National Natural Science
Foundation of China under Grant Nos. 90303004, 10533010, 19925523
and 10405029, and also in part by the Scientific Research Fund of
GUCAS(NO.055101BM03).

\end{document}